\documentclass{article}

\PassOptionsToPackage{numbers, compress}{natbib}

     \usepackage[preprint]{neurips_2020}

\usepackage[utf8]{inputenc} 
\usepackage[T1]{fontenc}    
\usepackage{hyperref}       
\usepackage{url}            
\usepackage{booktabs}       
\usepackage{amsfonts}       
\usepackage{nicefrac}       
\usepackage{microtype}      

\usepackage{amsmath}  
\usepackage{amssymb}
\usepackage[pdftex]{graphicx}

\newcommand{\vect}[1]{\boldsymbol{#1}}

\DeclareMathOperator{\tr}{tr}
\newcommand{\norm}[1]{\left\lVert#1\right\rVert}
\title{Using noise to probe recurrent neural network structure and prune synapses}

\author{
  Eli Moore \\
  Department of Mathematics\\
  University of California, Davis\\
  Davis, CA 95616 \\
  \texttt{elimoore@ucdavis.edu} \\
  \And
  Rishidev Chaudhuri\\
  Center for Neuroscience \\
  Department of Mathematics\\
  Department of Neurobiology, Physiology and Behavior\\
  University of California, Davis\\
  Davis, CA 95616\\
  \texttt{rchaudhuri@ucdavis.edu} \\
}

\begin{document}
\maketitle

\begin{abstract}
Many networks in the brain are sparsely connected, and the brain eliminates synapses during development and learning. How could the brain decide which synapses to prune? In a recurrent network, determining the importance of a synapse between two neurons is a difficult computational problem, depending on the role that both neurons play and on all possible pathways of information flow between them. Noise is ubiquitous in neural systems, and often considered an irritant to be overcome. Here we suggest that noise could play a functional role in synaptic pruning, allowing the brain to probe network structure and determine which synapses are redundant. We construct a simple, local, unsupervised plasticity rule that either strengthens or prunes synapses using only synaptic weight and the noise-driven covariance of the neighboring neurons. For a subset of linear and rectified-linear networks, we prove that this rule preserves the spectrum of the original matrix and hence preserves network dynamics even when the fraction of pruned synapses asymptotically approaches 1. The plasticity rule is biologically-plausible and may suggest a new role for noise in neural computation. 
\end{abstract}

\section{Introduction}
The brain eliminates synapses, dramatically during development but then across the lifespan \cite{kasai10, petanjek11, stein19}. The degree of synaptic pruning post-learning correlates with learning performance, suggesting an important functional role \cite{yang09, lai12}. Moreover, connection density is disrupted across a spectrum of diseases \cite{paus08, wass11, pievani14, heuvel14}. Determining how the brain finds and maintains sparse network structure is important to understand the brain's remarkable energy efficiency and replicate it in artificial neural networks as well as to understand changes in connection density with aging \cite{sowell03, damoiseaux17} and in disease \cite{paus08, wass11, pievani14, heuvel14}. 

In a highly-recurrent network with multiple pathways of information flow, it is difficult to determine which synapses are redundant and can be safely pruned, and which are important and should be retained. For example, even if a synapse between two neurons is strong, if information can travel between the neurons by alternative pathways then the synapse is redundant and can be removed. A biologically-plausible pruning rule must determine this higher-order structure using information locally available at the synapse.

Neural systems seem noisy, at multiple levels: neural activity contains large background fluctuations, responses to the same stimulus can be quite variable, and synapses often fail to propagate a signal \cite{softky93, koch04, faisal08, destexhe12}. Here we show that noise could play a useful computational role in synaptic pruning. Specifically, the pattern of activity correlations in a noise-driven network reflects higher-order network structure in exactly the form needed for good synaptic pruning (as predicted by a theoretical argument). We construct a local plasticity rule that either strengthens or prunes synapses with a probability given by the synaptic weight and the noise-driven covariance of the neighboring neurons. The plasticity rule is unsupervised and task-agnostic, seeking only to preserve existing network dynamics, whatever they are. Thus, it could act alongside learning or during separate pruning epochs (e.g., sleep), and does not restrict the learning rule in any way. 
 
We prove that, for a class of undirected linear and rectified linear networks, the pruning rule preserves multiple useful properties of the original network (including the spectrum and resting-state variances), even when the fraction of removed synapses approaches 1. The theoretical results link neural network pruning and noise-driven dynamical systems to a powerful body of results in sampling-based graph sparsification \cite{spielman11, spielman_teng11, batson13} and to matrix concentration of measure tools \cite{rudelson99, ahlswede02, rudelson07, tropp12}.

\section{Problem setup}
We primarily consider linear neural networks of the form
\begin{align}
\frac{d\vect{x}}{dt} = -D\vect{x} + W\vect{x} + \vect{b}(t)  = Ax + \vect{b}(t)
\label{eq:lin_dynamics}
\end{align}
The vector $\vect{x}$ represents the firing rate of $N$ neurons, with $x_i$ the firing rate of the $i$-th neuron. $\vect{b}(t)$ is the external input to the neurons (including biases). $W$ is the matrix of weighted connections between the neurons, with $W_{ij}$ the connection strength from the $j$-th to the $i$-th neuron. $D$ is a diagonal matrix representing the intrinsic leak of activity (or the excitability of the neuron). Finally we define the matrix $A = -D + W$. We discuss generalizations to rectified linear networks in Section \ref{sec:gen_ex}.

The pruning rule seeks to generate a sparse network with corresponding matrix $A^{sparse}$ with two properties. First, the number of edges in the network (i.e., number of non-zero entries in $A^{sparse}$) should be small when compared to the $\sim N^2$ possible edges in the original network. Second, the dynamics of the pruned network
\begin{equation}
\frac{d\vect{x}}{dt} = A^{sparse}\vect{x} + \vect{b}(t)
\label{eq:pruned_lin_dynamics}
\end{equation}         
should be similar to the dynamics of the original network in Eq. \ref{eq:lin_dynamics}.

To measure the similarity of $A$ and $A^{sparse}$, we adopt the notion of spectral similarity \cite{spielman11, spielman_teng11} from the field of graph sparsification and require that for some small $\epsilon > 0$,
\begin{equation}\label{eq:spectral_approx}
|\vect{x}^T(A^{sparse} - A)\vect{x}| \leq \epsilon|\vect{x}^TA\vect{x}| \quad \forall \vect{x} \in \mathbb{R}^N.
\end{equation}
This notion of similarity is quite strong. For symmetric matrices it requires that the eigenvalues of $A^{sparse}$ approximate the eigenvalues of $A$ (and hence all the timescales of the resulting dynamics) to within a multiplicative factor $\epsilon$ (see SI S1.3). This closeness is much stronger than low rank approximation, which preserves only the largest eigenvalues. The pruning rule also approximately preserves matrix-vector products and eigenvectors corresponding to separated eigenvalues. The timescales and activity patterns of the dynamical system in Eq.  \ref{eq:lin_dynamics} are determined by the spectrum of $A$, and thus spectrum-preserving sparsification will (approximately) preserve dynamics.

\section{An unsupervised noise-driven anti-Hebbian pruning rule}
Consider the network in Eq. \ref{eq:lin_dynamics} when driven by independent noise at each node. We set $\vect{b}(t) = \vect{b} + \sigma\vect{\xi}(t)$ where $\vect{b}$ is an arbitrary vector of constant background input to the network, $\vect{\xi}$ is a vector of IID unit variance Gaussian white noise, and $\sigma$ is the standard deviation of the input noise. Let $C$ be the covariance matrix of the firing rates in response to this white noise input. For the synapse from neuron $j$ to neuron $i$, with weight $w_{ij}$, define the probability
\begin{equation}\label{eq:pruning_rule}
p_{ij} =
\begin{cases}
Kw_{ij}\left(C_{ii} + C_{jj} - 2C_{ij}\right) & \text{for } w_{ij}>0 \quad \text{(excitatory)} \\
K|w_{ij}|\left(C_{ii} + C_{jj} + 2C_{ij}\right) & \text{for } w_{ij}<0 \quad \text{(inhibitory)}.
\end{cases}
\end{equation}
Here $C_{ii}$ and $C_{jj}$ are the variances of the $i$th and $j$th neurons, and $C_{ij}$ is their covariance. $K$ is a proportionality constant and determines the density of the pruned network, which will have $NK/2$ total connections on average and thus average degree of $K/2$ per neuron (for unit variance noise and symmetric networks). 

Now consider a pruning process that independently preserves each edge with probability $p_{ij}$ yielding $A^{sparse}$, where for $i \neq j$,
\begin{equation}\label{eq:sampling_edges}
A^{sparse}_{ij} = 
\begin{cases}
A_{ij}/p_{ij} & \text{with probability } p_{ij} \\
0 & \text{otherwise}.
\end{cases}
\end{equation}
For the diagonal terms (i.e., leak / excitability) $A^{sparse}_{ii}$ we either preserve the original diagonal and set $A^{sparse}_{ii} = A_{ii}$ or define the perturbation $\Delta_i = \sum_{j \neq i} |A_{ij}^{sparse}| - \sum_{j \neq i} |A_{ij}|$ to be the change in total input to neuron $i$ and set $A^{sparse}_{ii} = A_{ii} - \Delta_i$. $\Delta_i$ is small with zero mean, and biologically corresponds to changing the excitability of neuron $i$ in response to a change in total input (excitability is known to be homeostatically regulated \cite{turrigiano17}). We call these the ``original diagonal'' and ``matched diagonal'' settings respectively. The proofs apply to the ``matched diagonal'' setting (empirically, similar results apply to the ``original diagonal'' setting). We will also refer to the pruning rule defined by Eqs. \ref{eq:pruning_rule}, \ref{eq:sampling_edges} (in both settings for the diagonal) as \textbf{noise-prune} going forward.

The noise-prune rule is predicted by a theoretical argument (see next section), but has an appealingly simple interpretation and we here provide some intuition for why it might work. First, note that the probability to preserve a synapse depends on the magnitude of its weight, $|w_{ij}|$. Thus, all else being equal, synapses with larger weight are more important and are preserved. The remainder of the expression for the preservation probability is $\left(C_{ii} + C_{jj} \pm 2C_{ij}\right)$, which we call the diff-cov term. This term can be slightly rewritten as $2\tilde{C}_{ij}(1 \pm C_{ij}/\tilde{C}_{ij})$, where $\tilde{C}_{ij} = (C_{ii} + C_{jj})/2$ is the mean variance of nodes $i$ and $j$. Thus, preservation probability is proportional to $\tilde{C}_{ij}$, reflecting that nodes with higher variance are considered more important and thus their connections are likely to be preserved (as in a PCA-like approximation). Finally, there is an anti-Hebbian term that for excitatory synapses takes the form $(1 - C_{ij}/\tilde{C}_{ij})$ (if $2\tilde{C}_{ij}$ is factored out) or $\left(C_{ii} + C_{jj} - 2C_{ij}\right)$. Synapses are thus likely to be preserved if they are weakly or anti-correlated despite having an excitatory connection. The equivalent term for inhibitory synapses is $\left(C_{ii} + C_{jj} + 2C_{ij}\right)$. The sign of the covariance is flipped, reflecting that inhibitory connections are expected to anti-correlate neurons. 

The covariance of neurons $i$ and $j$ depends both on the strength of the direct connection between them (i.e., $w_{ij}$) and on indirect connections through the rest of the network. Neurons that are highly correlated are likely to have multiple indirect connections, suggesting that the direct connection is redundant and can be pruned (see schematics in Fig. 1a,b). More generally, the pruning rule can be understood as probing whether neurons are more correlated than expected given the weight of their direct connection. If they are, the connection is likely to be redundant. 

For a simple example of why higher correlations indicate a connection that can be pruned, consider probing one's indirect connections to a friend by spreading a rumor about them and measuring how distorted the rumor is by the time they hear it (or, alternatively,  playing the children's game ``telephone''). More distortion (i.e., lower correlation) indicates that there are few indirect routes for information to flow through and thus the connection is likely to be important. Conversely, the less distorted the message, the more redundant the connection.

\begin{figure}
    \begin{center}
            \includegraphics[width=\linewidth]{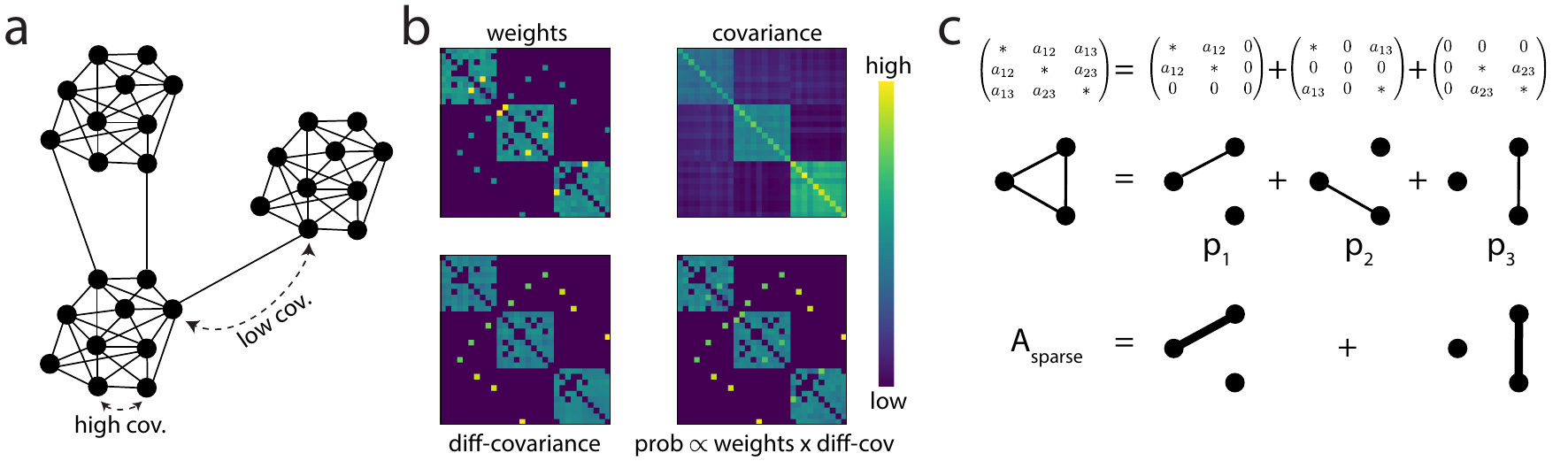}
    \end{center}
    \caption{{\bf A noise-driven unsupervised synaptic pruning rule.} (a) Schematic of a network where noisy fluctuations reflect higher-order connectivity structure. Network has 3 densely-connected clusters, with a few long-range connections. Covariance between neurons within a cluster is high compared to neurons participating in different clusters. (b) Pruning rule uses weights and covariances to identify important synapses. Top left: connection weights in a network with 3 densely-connected clusters and sparse connections between clusters. Also note the presence of a few strong connections within each cluster. Top right: covariance when driven by noise. Bottom left: difference of covariances (as in Eq. \ref{eq:pruning_rule}). Bottom right: sampling probabilities from pruning rule. The rule correctly identifies that the sparse connections between clusters are important and assigns them higher probability, along with the handful of exceptionally strong connections within a cluster. Most connections within a cluster are redundant and given lower probability. (c) Schematic of proof strategy. The original network (shown as a matrix in the top row and as a graph in the middle row) can be written as a sum of edge pieces. The edges are assigned sampling probabilities ($p_1$, $p_2$, $p_3$) that depend on weight and covariance. A given application of noise-prune yields a sparse network (bottom row) where some connections are preserved and strengthened (first and third edges) and others are pruned (second edge). For appropriate probabilities, the spectrum of $A^{sparse}$ is close to that of the original network.}
\label{fig:noise_prune_schematic}
\end{figure}

\section{Proofs}\label{sec:proofs}
We derive this pruning rule from a two-part theoretical argument. First, we consider a sampling-based approach to pruning that independently strengthens or removes each edge of a network with some probability (as in Eq. \ref{eq:sampling_edges}) and derive sampling probabilities that preserve network dynamics. The structure of the argument follows Spielman \& Srivastava (2011), with slight extension to signed symmetric diagonally-dominant neural network matrices. Second, we show that these theoretically-derived probabilities have a surprisingly simple expression in terms of the covariance of the network activity when driven by noisy fluctuations. Thus, there exists a simple and biologically-plausible way for neural networks to compute the sampling probabilities using local information. 

Note that the proof, but not the noise-prune rule itself, requires the matrix $A$ to be symmetric (corresponding to an undirected graph) and diagonally-dominant (corresponding to quite leaky neurons). These are strong restrictions that do not typically apply to neural networks, and we discuss generalizations and limitations later, including preliminary empirical results that show that noise-prune can work well even when these restrictions do not hold (see Fig. 2). 

\subsection{Derivation of probabilities}
Assume that the matrix $A$ from Eq. \ref{eq:lin_dynamics} is symmetric ($A_{ij} = A_{ji}$) and diagonally-dominant, meaning that $|A_{ii}| \geq \sum_{j \neq i} |A_{ij}| = \sum_{j \neq i} |w_{ij}|$. The diagonal entries of $A$ are negative, reflecting the leak, and thus $A$ is negative definite (note that eigenvalues must be negative for the linear system to be stable, but the argument can be extended to non-invertible matrices by working in the subspace orthogonal to the null space \cite{spielman11}). For notational convenience, we define the positive definite matrix $B = -A$ and consider $B$ instead of $A$ in this section. 

Given an edge $(i,j)$, $i > j$, with weight $w_{ij}$, define the edge matrix $X^{(i,j)}$ to have $i$th and $j$th diagonal entries $X^{(i,j)}_{ii} = X^{(i,j)}_{jj} = |w_{ij}|$. Set the $(i,j)$th and $(j,i)$th off-diagonal entries $X^{(i,j)}_{ij} = X^{(i,j)}_{ji} = -w_{ij}$ and remaining entries $0$ (Fig. 1c for a schematic). Thus, $X^{(i,j)}$ has off-diagonal pieces equal to negative edge weight and diagonal pieces equal to its magnitude. Also define $X_{ii}^{(i,i)} = B_{ii} - \sum_{j \neq i} |w_{ij}|$, with remaining entries $0$. Because $B$ is diagonally-dominant with positive diagonal, the single non-zero entry of $X^{(i,i)}$ is positive. For simplicity, we consider matrices where $X^{(i,i)} = 0$, but it is straightforward to include non-zero $X^{(i,i)}$ (SI S1.1). $B$ can be written as a sum over edge matrices as $B = \sum_{i > j} X^{(i,j)}$.

Now define the random matrix $\tilde{X}^{ij}$ as
\begin{equation}
\tilde{X}^{ij} = 
\begin{cases}
X^{(i,j)}/p_{ij} &\text{ with probability } p_{ij} \\
0 &\text{ otherwise}
\end{cases}
\end{equation}
And define $B^{sparse} = \sum_{i> j } \tilde{X}^{ij}$. For any choice of $p_{ij}$, $\mathbb{E}[B^{sparse}] = B$. Thus on average $B^{sparse}$ is the original matrix.  Also note that the average number of edges in $B^{sparse}$, $\mathbb{E}[N_{edges}] = \sum_{i>j} p_{ij}$. 

If the $p_{ij}$ are close to $1$, then most edges will be included in any realization of $B^{sparse}$ and it will be close to $B$, but not sparse. If the $p_{ij}$ are small, then $B^{sparse}$ will be sparse but might be a poor approximation to $B$. A good algorithm will choose the $p_{ij}$'s to ensure both that $B^{sparse}$ is close to $B$ (in some appropriate sense) and that the number of non-zero edges is small. 

To determine good sampling probabilities, we follow Spielman \& Srivastava (2011) and first transform $B$ to the identity matrix. Note that $I = B^{-1/2}BB^{-1/2}$, where $I$ is the identity matrix and $B^{-1/2}$ is the matrix that squares to $B^{-1}$ (well-defined because $B$ is symmetric positive definite). Define $\tilde{Y}^{ij} =  B^{-1/2}\tilde{X}^{ij}B^{-1/2}$ and $\tilde{I} = \sum_{i>j} \tilde{Y}^{ij} = B^{-1/2}B^{sparse}B^{-1/2}$. Note that $\mathbb{E}[\tilde{I}] = I$.

The matrix Chernoff bound \cite{rudelson99, ahlswede02, rudelson07, tropp12} bounds the probability that $\tilde{I}$ is far from $I$. Let $M$ be an upper bound on the $\tilde{Y}^{(i,j)}$'s, so that $0 \leq ||\tilde{Y}^{(i,j)}||_2 \leq M$. Let $\tilde{\lambda}_{min}$ and $\tilde{\lambda}_{max}$ be the minimum and maximum eigenvalues of $\tilde{I}$. For given $0 < \epsilon < 1$, the bound guarantees that
\begin{equation}\label{eq:matrix_chernoff_identity_mat}
P\left[\tilde{\lambda}_{min} \leq (1-\epsilon) \right]  \leq N \left(e^{-\epsilon^2/2}\right)^{1/M} \,\,\text{and}\,\,
P\left[\tilde{\lambda}_{max} \geq (1+\epsilon) \right]  \leq N \left(e^{-\epsilon^2/3}\right)^{1/M}
\end{equation}
A good approximation thus requires that $M$ be small. On the other hand, since the sampled pieces are rescaled by $1/p_{ij}$, a sparser approximation (smaller $p_{ij}$) corresponds to larger $M$.

For each $(i,j)$, the maximum value that $||\tilde{Y}^{(i,j)}||_2$ takes is $\frac{1}{p_{ij}}||B^{-1/2}X^{(i,j)}B^{-1/2}||$. Set 
\begin{equation}\label{eq:prob_B_defn}
\frac{p_{ij}}{K_{deg}} = ||B^{-1/2}X^{(i,j)}B^{-1/2}|| =  \tr(B^{-1}X^{(i,j)}) = |w_{ij}|(B^{-1}_{ii} + B^{-1}_{jj} - \text{sign}(w_{ij}) 2B^{-1}_{ij}),
\end{equation}
for some constant $K_{deg}$, where the second equality holds since the trace is cyclic and equal to the $2$-norm of a rank-$1$ positive semi-definite matrix. This equalizes the maximum value across $\tilde{Y}^{(i,j)}$, yielding $M = 1/K_{deg}$. 

For any given $\epsilon$, ensuring that the probabilities in Eq. \ref{eq:matrix_chernoff_identity_mat} are small requires that $K_{deg} \geq 4\log(N)/\epsilon^2$ (the constant 4 is chosen semi-arbitrarily to ensure small probability for reasonable $N$ and other values $>3$ can be chosen). Thus $K_{deg} = 4\log(N)/\epsilon^2$ guarantees that the eigenvalues of $\tilde{I}$ lie within $[1-\epsilon, 1+\epsilon]$ with high probability (w.h.p.). Consequently, w.h.p., we have
\begin{equation}
(1-\epsilon)\vect{y}^T\vect{y} \leq \vect{y}^T\tilde{I}\vect{y} \leq (1+\epsilon)\vect{y}^T\vect{y} \quad \forall \vect{y} \in \mathbb{R}^N.
\label{eq:spectral_approx_id}
\end{equation}
Given some $\vect{x} \in \mathbb{R}^N$, set $\vect{y} = B^{1/2}\vect{x}$ yielding that w.h.p.,
\begin{equation}\label{eq:spectral_approx_B}
(1-\epsilon)\vect{x}^TB\vect{x} \leq \vect{x}^TB^{sparse}\vect{x} \leq (1+\epsilon)\vect{x}^TB\vect{x} \quad \forall \vect{x} \in \mathbb{R}^N.
\end{equation}
And observing that $B = -A$ yields the desired approximation.

The average number of edges in the pruned network $\langle N_{edges} \rangle = \sum_{i >j} p_{ij}$ (and a standard scalar Chernoff bound shows that fluctuations around the mean are small). Note that $\sum_{i > j} ||B^{-1/2}X^{(i,j)}B^{-1/2}|| = N$ (proof in SI S1.1). Hence $\langle N_{edges} \rangle = \sum_{i >j} p_{ij} = NK_{deg}$. Consequently, if $K_{deg}=4\log(N)/\epsilon^2$ then, in terms of $\epsilon$, $\langle N_{edges} \rangle = 4N\log(N)/\epsilon^2$.

As with the sparsification of graph Laplacians \cite{spielman11}, for a fixed relative approximation ($\epsilon$) to $A$, the number of edges in $A^{sparse}$ need only be $O(N\log(N))$. This is very strong: if the original network is dense then it has $\sim N^2$ edges; thus the fraction of edges needed for fixed $\epsilon$ goes to $0$ with increasing $N$. On the other hand, if the number of edges in $A^{sparse}$ is a small but non-vanishing fraction of the edges in $A$, then the approximation becomes arbitrarily good with increasing $N$ (i.e., $\epsilon \to 0$).

\subsection{Probabilities from noise-driven covariance}
Consider the network of Eq. \ref{eq:lin_dynamics} when driven by uncorrelated white noise of variance $\sigma^2$ at each node. Set the constant background input $\vect{b} = 0$ for simplicity (this just shifts the mean to $0$). The covariance matrix, $C$ of the resulting dynamics is $C = \mathbb{E}[\vect{x}\vect{x}^T]$ and satisfies the Lyapunov equation \cite{gardiner85, trentelman12}:
\begin{equation}
AC + CA^* = -\sigma^2I.
\end{equation}

Let $A$ be a normal matrix (meaning $AA^* = A^*A$, where $A^*$ is the conjugate transpose of $A$; this category includes symmetric matrices, such as the ones we consider). Define $A_{symm} = (A + A^*)/2$. It is straightforward to show that $C \propto A_{symm}^{-1}$ (see SI S1.2 for details). In particular, if $A$ is symmetric then $C = -\sigma^2A^{-1}/2$. Substituting $2C/\sigma^2$ for $B^{-1} = -A^{-1}$ in Eq. \ref{eq:prob_B_defn} yields
\begin{equation}\label{eq:samp_prob_cov}
p_{ij} = K |w_{ij}|(C_{ii} + C_{jj} - \text{sign}(w_{ij}) 2C_{ij}),
\end{equation}
with $K = 2K_{deg}/\sigma^2$. Thus, perhaps surprisingly, the pattern of noise-driven correlations exactly encodes the optimal sampling probabilities predicted by the matrix Chernoff bound.

\section{Numerical results}
\begin{figure}
    \begin{center}
            \includegraphics[width=\linewidth]{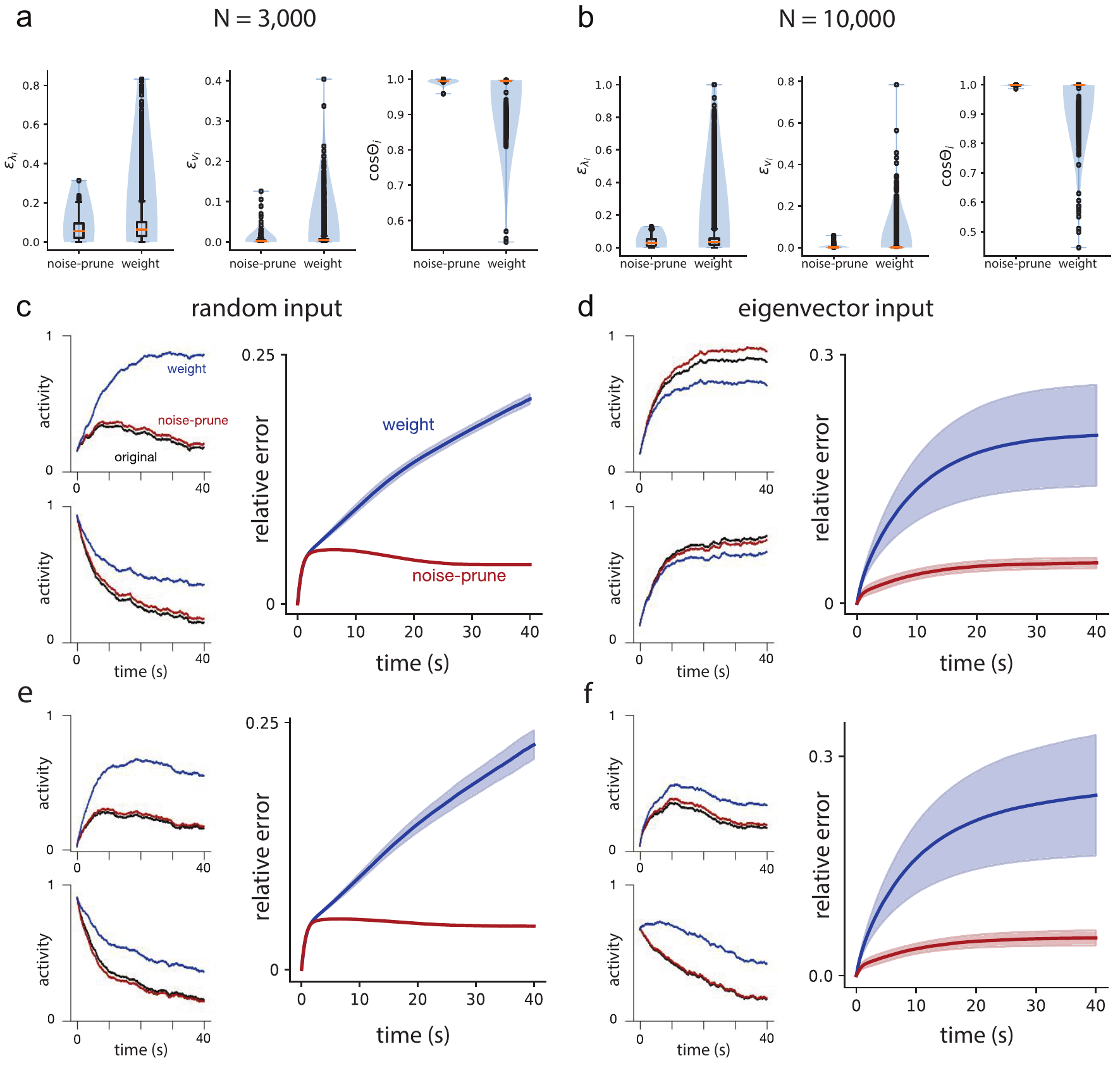}
    \end{center}
    \caption{{\bf Noise-prune performance on clustered symmetric and non-symmetric networks.}
(a) Performance of noise-prune (left box in each panel) and weight-based pruning (right box in each panel) on networks pruned to $10\%$ density. The left network of size $N = 3,000$ contains $3$ clusters of size $100$ and 1 cluster of size $2700$, with dense within-cluster connections ($60\%, \sim N(1,1)$) and sparse long-range connections ($5000 \text{ total}, \sim U(0,1)$). From left to right, panels show distribution of $\epsilon_{\lambda_i}$, $\epsilon_{v_i}$ and $\cos(\theta_i)$ (defined in text). Note that good performance corresponds to $\epsilon_{\lambda_i}$ and $\epsilon_{v_i}$ near $0$ and $\cos(\theta_i)$ near 1. Boxes show upper and lower quartiles, filled circles show outliers, violin plots show density estimate. (b) As in (a) but for larger clustered network ($N=10,000$, contains $10$ clusters of size $100$ and 1 cluster of size $9000$). (c-f) Dynamical response for networks with three clusters ($1000, 200,$ and $800$ nodes; connections distributed as in (a)). Black traces are the original unpruned network; red traces are networks pruned to $20\%$ sparsity with noise-prune in the matched diagonal setting; blue traces are networks pruned to $20\%$ sparsity using probabilities depending solely on weights. (c) Response of symmetric clustered network to random inputs. Panel shows trajectories from dynamical system $\frac{d\vect{x}}{dt} = A\vect{x} + \vect{b} + \vect{\xi}(t)$, where $\vect{b}$ is a small constant background input ($0.0002$), $\vect{\xi}(t)$ is gaussian white noise, and the initial condition $\vect{x}(0)$ is chosen with uniformly random entries $U(0,1)$. Left: response of two sample neurons for the three conditions. Right: mean (lines) and standard deviation (shaded area) of relative errors $||\vect{x}_{orig}(t) - \vect{x}_{np}(t)||_2/||\vect{x}_{orig}(t)||_2$ (red) and  $||\vect{x}_{orig}(t) - \vect{x}_{w}(t)||_2/||\vect{x}_{orig}(t)||_2$ (blue) over 20 different initial conditions and pruning runs. Here $\vect{x}_{orig},\vect{x}_{np},\vect{x}_{w}$ are dynamical responses of the original, noise-pruned, and weight-pruned network respectively. (d) As in (c), but with $\vect{b} = \vect{x}(0) = \vect{v}_i$ where $\vect{v}_i$ is the eigenvector corresponding to the $i$th largest eigenvalue of $A$ (or, equivalently, the $i$th smallest eigenvalue of $B$). Results averaged over $20$ slowest eigenvectors ($i = 1, \ldots, 20$). (e), (f) Analogous to (c), (d) respectively but for networks with non-symmetric connections.}
    \label{fig:numerical_results}
\end{figure}

In Fig. \ref{fig:numerical_results} we show the performance of noise-prune (in the matched diagonal regime) on diagonally-dominant networks with clustered structure (parameters in figure caption). We compare it to a control case in which edges are sampled and either strengthened or pruned (as in Eq. \ref{eq:sampling_edges}) but with probabilities just proportional to weight (i.e., without a covariance term and thus without accounting for higher-order network structure). The proportionality constant for the control is chosen to match the expected number of edges preserved by noise-prune.

The box plots in the first columns of Fig. \ref{fig:numerical_results}a,b show the distribution of relative change in eigenvalues of the pruned network when compared to the original network, given by $\epsilon_{\lambda_i} =  \left|\frac{\tilde{\lambda}_i}{\lambda_i} - 1\right|$, where $\tilde \lambda_i$ is the $i$th eigenvalue of $A^{sparse}$, and $\lambda_i$ is the $i$th eigenvalue of $A$. The box plots in the second column compare the relative change in quadratic forms $\epsilon_{v_i} =  \left|\frac{v_i^T A^{sparse}v_i}{v_i^T Av_i} - 1 \right|= \left|\frac{v_i^T A^{sparse}v_i}{\lambda_i} - 1\right|$ for the two approximations, where $(v_i,\lambda_i)$ is the $i$th eigenvector-eigenvalue pair of $A$. Lastly, the box plots in the third column measure how close the eigenvectors of the original network are to being eigenvectors of the pruned network using the normalized dot products of the eigenvectors before and after applying $A^{sparse}$: $\cos(\theta_i) = \frac{\left|v_i^T A^{sparse}v_i\right|}{\norm{A^{sparse}v_i}}$. In all cases, noise-prune performs better than the control, with the performance improving as the networks get larger (panel a vs. b).

We also compare the dynamical response of networks to various inputs before and after pruning.  In Fig. \ref{fig:numerical_results}c we show the response of symmetric clustered networks to random inputs before and after pruning, and find that noise-prune preserves both the responses of individual nodes (left panel) and the network response trajectory as a whole (right panel). We also find similar preservation for structured inputs directed along the slow eigenvectors of the network coupling matrix, which reflect integrative shared dynamical modes that may be used for computation, Fig. \ref{fig:numerical_results}d. Moreover, noise-prune significantly outperforms the purely weight-based strategy (red vs. blue) and thus using the higher-order structure reflected in the noise covariances dramatically improves the preservation of dynamics in the pruned network.

The theoretical results apply to the case of symmetric matrices but the pruning rule itself is quite general. We thus empirically characterize noise-prune on non-symmetric clustered networks for both random and eigenvector inputs, Fig. \ref{fig:numerical_results}e and f. Again, noise-prune preserves network dynamics and does much better than a control strategy that relies only on weight, suggesting that good performance extends beyond the theoretical guarantees.

\section{Extensions}\label{sec:gen_ex}
We next briefly describe some extensions of the framework described above (further details in SI).
 
\paragraph{Approximate probabilities}
The pruning is robust to approximate probabilities (as with graph Laplacian sparsification \cite{spielman11}). To see robustness, note that the probabilities (a) determine the upper bound $M$ used in Eq. \ref{eq:matrix_chernoff_identity_mat} and (b) determine $\langle N_{edges} \rangle$ through their sum. Consequently, if some edges are undersampled by a multiplicative factor $\alpha<1$ (i.e., probabilities $\hat{p}_{ij} = \alpha p_{ij}$ where the $p_{ij}$'s are the probabilities in Eq. \ref{eq:samp_prob_cov}) then the bound $M$ will be inflated by a factor of $1/\alpha$ and Eq. \ref{eq:spectral_approx_B} will still hold albeit with a larger $\hat{\epsilon} = \epsilon/\sqrt{\alpha}$, while the pruned network will have fewer edges. Moreover, sampling some edges with a probability higher than the $p_{ij}$'s will not harm the bound in Eq. \ref{eq:spectral_approx_B} (and will simply increase the number of preserved edges linearly in the degree of oversampling). In particular, any subset of the probabilities can be set to $1$; thus the pruning rule can be naturally applied only to a subset of connections. For more details on these arguments see SI S1.4.

\paragraph{Near-diagonally dominant networks}
Given a matrix $A$ with eigenvalues $\lambda_i$ and some constant $\gamma$, note that the matrix $A_\gamma = A + \gamma I$ has eigenvalues $\lambda_i + \gamma$ and the same eigenvectors as $A$. If $A$ is not diagonally-dominant, the application of noise-prune to $A$ can be analyzed by considering its effect on $A_\gamma$, with $\gamma$ chosen large enough that $A_\gamma$ is diagonally-dominant. There are two additional sources of error in the analysis: first, the probabilities are derived from the covariance matrix of $A$ and are thus sub-optimal for $A_\gamma$; second, the approximation of Eq. \ref{eq:spectral_approx_B} holds for $A_\gamma$ with some $\epsilon$ and the corresponding equation for $A$ includes an additive term of magnitude $\epsilon\gamma$ (see SI S2.1 for details).

 \paragraph{Rectified linear units}
Let $[\cdot]_{+}=\max[0,\cdot]$ be a rectified linear activation function and consider the recurrent neural network 
\begin{align}\label{eq:relu_dynamics}
\frac{d\vect{x}}{dt} = -D\vect{x} + \left[W\vect{x} + \vect{b}(t)\right]_+. 
\end{align}
As before, define $A = -D + W$. Let $A^{sparse}$ be the result of applying noise-prune to $A$ using the probabilities from the linear network defined by $A$ (consequently Eq. \ref{eq:spectral_approx} holds for $A$, $A^{sparse})$.

Let $\Gamma(t) = \{i : \sum_j W_{ij}x_j + b_j(t) >0\}$ be the indices of neurons that receive suprathreshold input at time $t$. Define $A_{\Gamma(t)}$ and $A^{sparse}_{\Gamma(t)}$ to be the submatrices produced by removing the rows and columns of $A$ and $A^{sparse}$ corresponding to indices not in $\Gamma$. The dynamics of the network in Eq. \ref{eq:relu_dynamics} is approximately determined by the set of linear systems with coupling matrices $A_{\Gamma(t)}$, $A^{sparse}_{\Gamma(t)}$ (proved in SI S2.2). Here, we note that the approximation Eq. \ref{eq:spectral_approx} for $A$, $A^{sparse}$ implies the same approximation for $A_{\Gamma(t)}$ and $A^{sparse}_{\Gamma(t)}$. Specifically, given some $\Gamma(t)$ with size $|\Gamma(t)|$, let $\Gamma(t,j)$ be the index of the $j$-th active neuron. Now given  $\vect{y} \in \mathbb{R}^{|\Gamma(t)|}$, define a corresponding $\vect{x} \in \mathbb{R}^N$ as $\vect{x}(\Gamma(t, j))=y(j)$ and remaining entries 0. Then $\vect{y}^T A^{sparse}_{\Gamma(t)}\vect{y} = \vect{x}^TA^{sparse}\vect{x}$, and similarly for $A_{\Gamma(t)}$ and $A$. Substituting into Eq. \ref{eq:spectral_approx}  shows that the approximation holds for $A_{\Gamma(t)}$, $A^{sparse}_{\Gamma(t)}$.

The argument requires sampling probabilities computed from the covariance matrix of the dynamical system with coupling matrix $A$. A simple way to determine these is to add non-specific background excitation or global fluctuations in excitability to the network to push neurons away from the threshold. Intriguingly, such global excitability fluctuations are observed during slow-wave sleep \cite{steriade93, vanhatalo04}.

\section{Discussion}
The structure of the sampling argument, the notion of spectral approximation, and the use of matrix concentration of measure tools are drawn from a rich body of work on graph sparsification \cite{spielman11, spielman_teng11, batson13}, particularly the beautiful paper of Spielman \& Srivastava (2011). Our study links these results with neural networks and noisy dynamical systems. In the graph Laplacian context, the counterpart of the diff-cov matrix (see Eq. \ref{eq:pruning_rule}) is ``effective resistance'', which measures the electrical resistance between nodes if the graph is considered a weighted resistor network. Effective resistance has multiple nice properties \cite{ghosh08, spielman11}, such as forming a natural metric \cite{klein93}, and the diff-cov matrix may be similarly useful for neural networks. Conversely, a difference of covariances has recently been suggested to generalize effective resistance to directed graphs \cite{young15}. There may be further useful connections to be drawn between this set of ideas and noise-driven dynamics in neural networks.

A number of studies have investigated task-dependent pruning of connections in artificial neural networks, often with very compelling results \cite{lecun90, reed93, hassibi93, han15, dong17, bellec18, frankle19, narang17, lee19, baykal19, blalock20}. Current state-of-the-art approaches in machine learning typically train a network to good performance on a task, assign a measure of importance to each connection in the network (often its weight and sometimes a measure of impact on the task cost function such as terms in the Hessian), remove connections from the network according to this importance measure, and then repeat the cycle of training and pruning (see \cite{blalock20} for a recent review). Such approaches have been extremely successful, yielding networks with greatly reduced density (as little as a few percent of the original) while preserving task performance. Our work is complementary to these pruning studies in three ways. First, these studies focus on the supervised, (typically) feedforward setting, and algorithms are not usually biologically plausible. By contrast, the current study seeks an unsupervised, biologically-plausible algorithm for recurrent networks. Second, most existing studies typically seek good empirical performance in quite challenging real-world applications rather than theoretical results, while we focus on developing strong theoretical results in a limited setting. And finally, existing algorithms that prune connections typically do so either based on connection weight or a nonlocal measure of cost function sensitivity, while we combine weight with a local term that extracts a connection's importance to the network from activity fluctuations. Our study is most reminiscent of unsupervised approaches that merge or remove highly correlated neurons \cite{srinivas15, mariet15, babaeizadeh16}, though the setting, algorithms and theoretical guarantees are quite different, and we consider weight pruning rather than removing entire neurons. Note that we do not expect noise-prune to be competitive with state-of-the-art supervised approaches in machine learning when measured by preserving performance on a given task (rather than preserving dynamics). However, the novel perspective provided by noise-prune and the theoretical results may be useful in developing more powerful algorithms for task-driven pruning.

The proofs apply to the limited case of symmetric diagonally-dominant linear and rectified linear recurrent networks. Certain networks in the brain may potentially be modeled as diagonally-dominant (e.g., in the high-conductance regime, when membrane time constants are very small \cite{destexhe03}), though it is unclear how good this approximation will be. More importantly, connections in biological neural networks are not symmetric. The framework may apply more naturally to excitatory (or inhibitory) sub-networks with a higher probability of reciprocal connectivity \cite{song05}, and especially to cell assemblies that code for the same stimulus or concept \cite{buzsaki10}. Finally, biological networks are nonlinear. Thus, the theoretical framework presented is far from general. 

However, we highlight two causes for optimism. First, in the limited regime where the theory applies, results are very strong and robust (as in graph Laplacian sparsification \cite{spielman11, spielman_teng11}), able to asymptotically preserve the entire spectrum even when the fraction of retained edges goes to 0. Preservation of the entire spectrum is likely too strong for neural networks, which often show redundant coding and low-dimensional dynamics. It may be possible to more weakly approximate a broader family of networks. Second, the noise-prune rule itself (Eq. \ref{eq:pruning_rule}) does not require particular network structure and can in principle be applied to any recurrent network (note that covariance for a general normal matrix is determined by the symmetric part of the matrix). Indeed, we empirically find that noise-prune preserves dynamics in non-symmetric clustered networks, Fig. \ref{fig:numerical_results}e, f, and thus shows good performance beyond the regime where theoretical guarantees hold. A more exhaustive empirical characterization of noise-prune is beyond the scope of the present study, but this is a natural direction for future work. 

The pruning rule uses randomness twice. First, it uses noisy fluctuations in activity to probe network structure and make global information locally available in the form of activity correlations between pairs of neurons. Second, it randomly decides whether to preserve and strengthen or prune a connection. This use of randomness is inspired by seemingly ubiquitous noise at multiple levels in neural systems \cite{softky93, koch04, faisal08, destexhe12}. It is still unclear how much of this ``noise'' reflects the encoding of unknown variables as opposed to genuine randomness, and to what degree noise is averaged away as opposed to being used as a computational resource. However, randomized algorithms are often appealingly simple, powerful and easy to parallelize, and it is plausible and widely speculated that brains have evolved to take computational advantage of biological noise \cite{faisal08, mcdonnell11}. 

Unlike pruning rules that remove (typically weak) synapses and simply preserve the others, the (subset of) synapses targeted by noise-prune are either removed or strengthened, reminiscent of observations that small spines on neurons are highly variable and liable to either vanish or grow and stabilize \cite{holtmaat05}. More generally, a strengthen-or-prune rule like that in Eq. \ref{eq:sampling_edges} can be applied with different sampling probabilities, which may be appropriate for different settings, and synapses can be strengthened or weakened rather than pruned. If weights and probabilities are chosen to preserve synaptic weights on average (which is a natural target for an unsupervised algorithm), then the approach approximately preserves total synaptic input to and output from a neuron as well as the dynamics resulting from a given input or network activity state. The theoretical approach may thus be more generally useful in settings where synaptic weight is redistributed across synapses (such as in some homeostatic mechanisms \cite{turrigiano17}). 

In this study we have focused on pruning synapses while preserving existing network dynamics, thus approaching pruning primarily as resource conservation. Pruning in the brain may serve other functions as well, such as making networks faster or more robust to noise. Given that the pruned network needs to carry out a similar set of input-output transformations to the original network, dynamical patterns are likely to be similar between unpruned and pruned networks and thus preservation of dynamics such as proposed here could be used as a building block to investigate more complex pruning algorithms that optimize other features of network responses.


The approach presented here suggests decomposing into two pieces the difficult problem of learning a sparse network solution to a task. First, a greedy task-driven learning epoch that adds synapses where they might be needed, regardless of efficiency (such as would be expected from correlational / Hebbian learning processes). And second, a noisy, task-agnostic, anti-Hebbian epoch during which a subset of synapses enter a labile state and are either consolidated or pruned. The second regime is reminiscent of theories of sleep \cite{tononi14,levenstein17} and it will be interesting to attempt to connect sleep phenomenology with the algorithm presented in this study.

\section*{Broader Impact}
While the larger question that motivates this study (how the brain might prune synapses) is of great practical interest, the results presented here are purely theoretical and quite abstract, and we do not foresee any immediate societal consequences or ethical issues.

\begin{ack}
We thank E. Natale, C. Papadimitriou and L. Turin for helpful discussions on graph sparsification and A. Bernacchia, K. Mimmack and E. Natale for comments on the manuscript. Part of this work was conceived when RC was a Google Research Fellow at the Simons Institute for the Theory of Computing at UC Berkeley. EM was partially supported by a UC Davis Summer GSR Award for Engineering or Computer-related Applications and Methods. The authors were benefited by participating in the activities of the UC Davis TETRAPODS Institute of Data Science, which has been funded by the NSF TRIPODS grant CCF-1934568.


\end{ack}

\bibliography{noisy_pruning}

%
%
%
%
%

\end{document}


\begin{center}
\textbf{\Huge Supplementary Information}
\end{center}

In this section we expand on the arguments in the main text. Note that, for completeness, some portions of the main text are repeated here.

\tableofcontents

\section{Theoretical framework underlying noise-prune}
\label{sec:basic_proof_results}
We consider the $N \times N$ coupling matrix of the linear system 
\begin{equation}\label{eq:lin_dynamics}
\frac{dx}{dt} = Ax + \vect{b}(t),
\end{equation}
and describe how to construct a sparse matrix $A^{sparse}$ whose spectrum (and hence dynamics) are similar to $A$. 

To measure the similarity of $A$ and $A^{sparse}$, we adopt the notion of spectral similarity \cite{spielman11, spielman_teng11} from the field of graph sparsification and require that for some small $\epsilon > 0$,
\begin{equation}\label{eq:spectral_approx}
|\vect{x}^T(A^{sparse} - A)\vect{x}| \leq \epsilon|\vect{x}^TA\vect{x}| \quad \forall \vect{x} \in \mathbb{R}^N.
\end{equation} 

The primary theoretical insights of this section are that (a) results on the sparsification of graph Laplacians \cite{spielman11, spielman_teng11} can be applied, with slight generalization, to pruning signed symmetric diagonally-dominant linear neural networks and (b) that the covariance matrix of the network when driven by noise provides appropriate pruning probabilities. We also discuss what properties of the original network are preserved after sparsifying the matrix $A$, as well as how these maintained properties are affected when the sampling probabilities are changed.

\subsection{Sparsification of symmetric, diagonally-dominant networks}\label{subsec:sparsification_of_sym_dd_networks}
In this section we show how to construct spectral sparsifiers of $A$. We follow the proof of \cite{spielman11}, with some adaptation. 

Let $A$ be the coupling matrix of a linear system, as in Eq. \ref{eq:lin_dynamics}. Note that in order for the linear system to be stable, all the eigenvalues of $A$ must have negative real part (and hence the matrix must be invertible). A non-invertible coupling matrix would correspond to a network with an unrealistically long (i.e., infinite) time-constant.

We impose the further restrictions that $A$ be a symmetric, \textit{diagonally-dominant} matrix; that is, $A_{ij} = A_{ji}$ and $|A_{ii}| \geq \sum_{j \neq i} |A_{ij}|$. In the main text, we focused on the case where this inequality was saturated (i.e., $|A_{ii}| = \sum_{j \neq i} |A_{ij}|$). Here, we expand the proof to include a \textit{strictly diagonally-dominant} $A$, thus satisfying $|A_{ij}| > \sum_{j \neq i} |A_{ij}|$ (note that the argument is essentially the same and also that both the original and matched diagonal cases of noise-prune simply preserve any excess weight along the diagonal).  The diagonal entries of $A$ reflect the intrinsic leak of activity and are negative. Combined with the strict diagonal-dominance requirement, the negative diagonal also implies that the eigenvalues of $A$ are negative, as can be seen from, e.g., a Gershgorin disk argument. Note that the diagonal dominance condition is stronger than the requirement of negative eigenvalues. In the event that $A$ satisfies $|A_{ii}| = \sum_{j \neq i} |A_{ij}|$ as in the main text, invertibility is no longer guaranteed by the Gershgorin disk argument but we assume invertibility based on the stability of the equivalent linear system. We can also relax the invertibility condition by considering the pseudoinverse of $A$ and working in the subspace orthogonal to the nullspace of $A$ (as done for graph Laplacians \cite{spielman11}). To sum up, $A$ is negative definite since it is symmetric with negative eigenvalues. 

For notational convenience, set $B = -A$ (and note that $B$ is positive definite). The non-zero off-diagonal entries $b_{ij} = -a_{ij} = -w_{ij}$ correspond to the connections in the network (note that throughout $w_{ij}$ refers to the weight in $A$, i.e., $-b_{ij}$; the argument can be rewritten without introducing $B$ at the cost of extra minus signs). 

\vspace{1mm} \textbf{Edge decomposition} For each of these undirected connections $(i,j)$ with $i > j$, we define the edge matrix $X^{(i,j)}$ by
\begin{equation}\label{eq:edge_mat_defn}
X_{k\ell}^{(i,j)} = \begin{cases}
|w_{ij}| & \text{if } (k,l) = (i,i) \text{ or } (j,j) \\
-w_{ij} & \text{if } (k,l) = (i,j) \text{ or } (j,i) \\
0 & \text{otherwise.}
\end{cases}
\end{equation}
Note that there is no restriction on the sign of $w_{ij}$. Also notice that $X^{(i,j)}$ can be written as $\vect{v}_{ij}\vect{v}_{ij}^T$ where $\vect{v}_{ij} \in \mathbb{R}^N$ has $i$th entry $\sqrt{|w_{ij}|}$ and $j$th entry $-\sgn(w_{ij})\sqrt{|w_{ij}|}$.  Thus $X^{(i,j)}$ a rank-$1$ matrix. Moreover, since the non-zero eigenvalue is positive, the matrix is positive semidefinite. Also note that specifying $i > j$ above is simply a manner of convention to not double-count connections in the symmetric matrix.

We also define the matrix $X^{(i,i)}$ for all $i$ to have only a single non-zero entry $X_{ii}^{(i,i)} = B_{ii} - \sum_{j \neq i} |w_{ij}|$. Because $B$ is diagonally-dominant with positive diagonal, the single non-zero entry of $X^{(i,i)}$ is positive, again implying that $X^{(i,i)}$ is rank-1 positive semidefinite. We include these diagonal pieces in the sampling argument for completeness but will usually simply treat them as fixed.

The original matrix $B$ is the sum of these edge matrices, $B = \sum_{i \geq j} X^{(i,j)}$ (where the notation $\sum_{i \geq j}$ sums over all existing edge pairs where $i \geq j$).

\vspace{1mm} \textbf{Sampling edges} Now, for $i \geq j$, define the random matrix $\tilde X^{ij}$ as 
\begin{equation}\label{eq:sampling_edges}
    \tilde X^{ij} = \begin{cases}
    X^{(i,j)}/p_{ij} & \text{with probability } p_{ij} \\
    0 & \text{otherwise,}
    \end{cases}
\end{equation}
where $0 < p_{ij}\leq 1$ is some probability we will determine below. Observe that, regardless of the choice of the $p_{ij}$'s, $\mathbb{E}[\tilde X^{ij}] = p_{ij}X^{(i,j)}/p_{ij} = X^{(i,j)}$. Correspondingly, for any set of probabilities, $\vect{p}$, we can define the matrix $B^{sparse, \vect{p}} = \sum_{i \geq j} \tilde X^{ij}$ and we have $\mathbb{E}[B^{sparse, \vect{p}}] = B$ (note that $B^{sparse, \vect{p}}$ will only be sparse if the $p_{ij}$ are small).

\vspace{1mm} \textbf{Transformation to identity}
Analogous to Spielman \& Srivastava (2011), we implement their argument in our framework by first transforming $B$ into the identity matrix $I$ and finding an appropriate approximation $\tilde I$, with the goal of transforming back and arriving at our desired sparsifier $B^{sparse}$. This step is crucial for preserving the entire spectrum (as required by Eq. \ref{eq:spectral_approx}), rather than only the largest eigenvalue (and leads to the diff-cov term in the probabilities). 

First observe that $I = B^{-1/2}BB^{-1/2}$, where $B^{-1/2}$ is the matrix whose square is $B^{-1}$ ($B^{-1/2}$ exists since $B$ is invertible and diagonalizable and moreover is real-valued since $B$ is positive definite).\footnote{If the eigenvector decomposition of $B$ is $UDU^{-1}$ then $B^{-1/2}$ can be constructed as $UD^{-1/2}U^{-1}$, where the entries of $D^{-1/2}$ are the inverse square roots of the corresponding entries of $D$).} Then, defining $Y^{(i,j)} = B^{-1/2}X^{(i,j)}B^{-1/2}$, we have
\begin{equation}
I = B^{-1/2}BB^{-1/2} = \sum_{i \geq j} B^{-1/2}X^{(i,j)}B^{-1/2} = \sum_{i \geq j}Y^{(i,j)}.
\label{eq:id_in_terms_of_B}
\end{equation}
This gives motivation to define the random matrices $\tilde Y^{ij}  = B^{-1/2}\tilde X^{ij} B^{-1/2}$ and $\tilde I = \sum_{i \geq j}\tilde Y^{ij}$. Note that $E(\tilde I) = I$.

Now, for given $0<\epsilon<1$, our goal will be to choose $p_{ij}$ in order to guarantee that
\begin{equation}
    \vect{y}^T(1-\epsilon) I\vect{y} \leq \vect{y}^T\tilde I\vect{y} \leq \vect{y}^T(1+\epsilon)I\vect{y} \quad \forall \vect{y} \in \mathbb{R}^N,
    \label{eq:id_psd_ineq}
\end{equation}
with high probability (w.h.p.). If we can do so, then for a given $\vect x \in \mathbb R^N$, we can set $\vect{y} = B^{1/2}\vect{x}$ in order to arrive at, w.h.p.,
\begin{equation}
    \vect{x}^T(1-\epsilon) B\vect{x} \leq \vect{x}^TB^{sparse}\vect{x} \leq \vect{x}^T(1+\epsilon)B\vect{x} \quad \forall \vect{x} \in \mathbb{R}^N,
    \label{eq:B_psd_ineq}
\end{equation}
where $B^{sparse} = B^{-1/2}\tilde I B^{-1/2} = \sum_{i \geq j} \tilde X^{ij}$, which provides the desired approximation.

\vspace{1mm} \textbf{Probabilities from matrix Chernoff bound} We want our $p_{ij}$ to be as small as possible while still maintaining the inequalities Eq. \ref{eq:id_psd_ineq}, \ref{eq:B_psd_ineq}. To derive good choices for the $p_{ij}$'s, we apply the matrix Chernoff bound  \cite{rudelson99, ahlswede02, rudelson07, tropp12} to bound the fluctuations of $\tilde I = \sum_{i \geq j}\tilde Y^{ij}$ around its expectation value, $I$. Let $M$ be an upper bound on the $\tilde{Y}^{ij}$'s, so that $0 \leq ||\tilde{Y}^{ij}||_2 \leq M$. Let $\lambda_{min}$ and $\lambda_{max}$ indicate minimum and maximum eigenvalues. The bound then guarantees that

\begin{align}
P\left[\lambda_{min}\left(\sum_{i\geq j} \tilde{Y}^{ij} \right) \leq (1-\epsilon)\right] &\leq N\left(\frac{e^{-\epsilon}}{(1-\epsilon)^{(1-\epsilon)}}\right)^{1/M} \leq Ne^{-\epsilon^2/2M} \quad \text{for } 0 < \epsilon <1, \nonumber \\
P\left[\lambda_{max}\left(\sum_{i \geq j} \tilde{Y}^{ij} \right) \geq (1+\epsilon)\right] &\leq N\left(\frac{e^{\epsilon}}{(1+\epsilon)^{(1+\epsilon)}}\right)^{1/M}\leq Ne^{-\epsilon^2/3M} \quad \text{for } 0 < \epsilon \label{eq:mat_chernoff}
\end{align}

The hypothesis of the bound requires the spectral norm of the $\tilde Y^{ij}$'s to be uniformly bounded across all edges; i.e., $||\tilde Y^{ij}|| \leq M$. Moreover $||\tilde Y^{ij}||$ depends on $1/p_{ij}$, so smaller probabilities lead to a larger bound $M$. Thus we choose the $p_{ij}$ in order to minimize $M$. 

Since $||\tilde Y^{ij}||$ is either $\frac{1}{p_{ij}} \norm{Y^{(i,j)}}$ or $0$, choose $p_{ij}$ to equalize the upper bound on $||\tilde Y^{ij}||$ across all $i \geq j$:
\begin{equation}\label{eq:prob_B_defn}
p_{ij} = K_{deg}\norm{Y^{(i,j)}} = K_{deg}||B^{-1/2}X^{(i,j)}B^{-1/2}||
\end{equation}
where $K_{deg}$ is some constant. This guarantees that $||\tilde Y^{ij}|| \leq M = 1/K_{deg}$. Thus, if we take $K_{deg} \geq 4 \log(N)/\epsilon^2$, the probabilities in Eq. \ref{eq:mat_chernoff} are guaranteed to be smaller than $1/N$ and $1/N^{1/3}$, respectively. Consequently, this choice of probabilities guarantees that Eqs. \ref{eq:id_psd_ineq}, \ref{eq:B_psd_ineq} are satisfied w.h.p., as desired.

Note that the constant $4$ is chosen somewhat arbitrarily here, with a larger constant corresponding to faster-decaying probabilities in Eq. \ref{eq:mat_chernoff} but also a larger number of edges expected to be sampled (since each $\tilde Y^{ij}$ is less likely to take on the value of $0$).

\vspace{1mm} \textbf{Bound on number of edges} Since edge $(i,j)$ is independently included with probability $p_{ij}$, the expected number of edges in the network is $\langle N_{edges}\rangle = \sum_{i > j} p_{ij}$ (note the strict inequality here, as $i = j$ does not correspond to edges, but rather the leak in neuronal activity). 

We have 
\begin{equation}
\sum_{i > j} p_{ij} \leq \sum_{i \geq j} p_{ij} =  K_{deg} \sum_{i \geq j} \norm{Y^{(i,j)}} = K_{deg} \sum_{i \geq j} \norm{B^{-1/2}X^{(i,j)}B^{-1/2}}
\end{equation}

Note that $Y^{(i,j)} = \vect{u}_{ij}\vect{u}_{ij}^T$, where $\vect{u}_{ij} = B^{-1/2}\vect{v}_{ij}$ and $\vect{v}_{ij}$ is the vector defined after Eq. \ref{eq:edge_mat_defn}. Consequently, $Y^{(i,j)}$ is rank-1 with positive eigenvalue and $\norm{Y^{(i,j)}} = \tr{Y^{(i,j)}}$. This yields
\begin{equation}
    \sum_{i \geq j} \norm{Y^{(i,j)}} = \sum_{i \geq j} \tr{Y^{(i,j)}} = \tr\left({B^{-1/2}\sum_{i \geq  j}X^{(i,j)}B^{-1/2}}\right) = \tr(B^{-1/2}BB^{-1/2}) = \tr(I) = N.
\end{equation}

Thus we have
\begin{equation}
    \langle N_{edges} \rangle = \sum_{i > j}p_{ij} = \sum_{i > j} K_{deg}\norm{Y^{(i,j)}} \leq  NK_{deg}.
\end{equation}
\vspace{1mm} \textbf{Simple expression for probabilities} Note that $||B^{-1/2}X^{(i,j)}B^{-1/2}|| = \tr\left(B^{-1/2}X^{(i,j)}B^{-1/2}\right) = \tr\left(B^{-1}X^{(i,j)}\right)$, again using the fact that the trace of a positive semi-definite rank-$1$ matrix is its spectral norm, and that the trace is cyclic (and $B^{-1/2}B^{-1/2} = B^{-1}$ by definition). The product $B^{-1}X^{(i,j)}$ has only two non-zero diagonal terms: its $i$th diagonal element is given by $|w_{ij}|B^{-1}_{ii} - w_{ij}B^{-1}_{ij}$ and its $j$th diagonal element is given by $-w_{ij}B^{-1}_{ji} + |w_{ij}|B^{-1}_{jj}$. Using the trivial decomposition $w_{ij} = \sgn{(w_{ij})}|w_{ij}|$ and adding these two diagonal elements together, we see that 
\begin{equation}
    p_{ij} = K_{deg}\tr\left(B^{-1}X^{(i,j)}\right) = K_{deg}|w_{ij}|(B^{-1}_{ii} + B^{-1}_{jj} - \sgn(w_{ij})2B^{-1}_{ij}),
\end{equation}
where we note that $B^{-1}_{ij} = B^{-1}_{ji}$, since the inverse of a symmetric matrix is symmetric.

Similarly, the $p_{ii}$ are observed to be
\begin{equation}
    p_{ii} = K_{deg}||B^{-1/2}X^{(i,i)}B^{-1/2}|| =K_{deg}\tr\left(B^{-1/2}X^{(i,i)}B^{-1/2}\right) =K_{deg}\tr\left(B^{-1}X^{(i,i)}\right)
\end{equation}
where we again use the cyclic property of the trace. Since the product $B^{-1}X^{(i,i)}$ has only the single non-zero diagonal element $B^{-1}_{ii}(B_{ii} - \sum_{j \neq i} |w_{ij}|)$, we arrive at the simple expression $p_{ii} = K_{deg}B_{ii}^{-1}\left(B_{ii} - \sum_{j \neq i} |w_{ij}|\right)$. Note that in practice we simply set this probability to $1$, but include it here for completeness.

Finally, recall that $B = -A$ and note that $A^{sparse} = -B^{sparse}$ is the outcome of the pruning applied to $A$. Substituting for $B$ in terms of $A$, the sampling probabilities are
\begin{equation}
p_{ij} = -K_{deg} |w_{ij}|(A^{-1}_{ii} + A^{-1}_{jj} - \text{sign}(w_{ij}) 2A^{-1}_{ij})
\label{eq:samp_prob_matrix_inv}
\end{equation}

\subsection{Sampling probabilities from noise-driven covariance}
The matrix inverse term $-A^{-1}$ in Eq. \ref{eq:samp_prob_matrix_inv} has a natural interpretation in terms of the covariance matrix of the corresponding linear dynamical system when driven by white noise. When the network is driven by noise, the dynamics are
\begin{equation}
\frac{dx}{dt} = A\vect{x} + \sigma\vect{\xi}(t),
\label{eq:noise_driven_dynamics}
\end{equation}
where $\vect{\xi}$ is unit variance Gaussian white-noise at each neuron and $\sigma$ is the standard deviation of the noise (note that this is a stochastic differential equation).

The covariance matrix of the resulting dynamics is given as the solution to the Lyapunov equation \cite{gardiner85, trentelman12}:
\begin{equation}
AC + CA^* = -\sigma^2I.
\label{eq:lyapunov_cov_wnoise}
\end{equation}

Assume that $A$ is normal, meaning that $A^*A = AA^*$, where $A^*$ is the conjugate transpose of $A$. Note that all symmetric matrices are normal. Since $A$ is normal it can be diagonalized as $A = U\Lambda U^*$, where $\Lambda$ is a diagonal matrix of eigenvalues and $U$ is unitary.

Substituting the decomposition of $A$ into Eq. \ref{eq:lyapunov_cov_wnoise} we have
\begin{align}
-\sigma^2 I &= U\Lambda U^*C + CU\Lambda^*U^*
\end{align}
so that multiplying this equation through by $U^*$ on the left and $U$ on the right and defining $\tilde{C} = U^*CU$, we arrive at
\begin{align}
-\sigma^2 I &= \Lambda U^*CU + U^*CU\Lambda^* = \Lambda \tilde C + \tilde C \Lambda^*.
\end{align}

Since $\Lambda$ is diagonal, the equation can be solved for the entries of $\tilde{C}$. $\tilde{C}$ is diagonal, with diagonal entries $\tilde{C}_{ii} = -\frac{\sigma^2}{\lambda_i + \lambda_i^*}$, where $\lambda_i$ and $\lambda_i^*$ are the $i$-th diagonal entries of $\Lambda$ and $\Lambda^*$ respectively (i.e., the $i$-th eigenvalue of $A$). By definition $C = U\tilde{C}U^*$ and thus $C$ has the same eigenvectors as $A$, with eigenvalues given by the diagonal entries of $\tilde{C}$. 

Define the symmetric part of $A$ to be $A_{symm} = \frac{1}{2} \left(A + A^*\right)$ and observe that this has eigenvalues $\frac{1}{2} \left(\lambda_i + \lambda_i^*\right)$. Thus, $C = -\frac{\sigma^2}{2} A_{symm}^{-1}$. In particular, for the symmetric matrices considered in the previous section, $C = -\frac{\sigma^2}{2} A^{-1}$. Substituting into the theoretically-derived form for the sampling rule and absorbing $\frac{\sigma^2}{2}$ into the overall constant yields

\begin{equation}
p_{ij} = K |w_{ij}|(C_{ii} + C_{jj} - \text{sign}(w_{ij}) 2C_{ij})
\label{eq:samp_prob_cov}
\end{equation}

\subsection{What is preserved}
The notion of spectral sparsification that we adopt from the graph Laplacian literature \cite{spielman_teng11, spielman11} (see Eq. \ref{eq:spectral_approx}) is quite strong and here we briefly discuss some of the properties it entails. 

Recall that, given $0 < \epsilon < 1$, Eq. \ref{eq:B_psd_ineq} guarantees that 
\begin{equation}
    \vect{x}^T(1-\epsilon) B\vect{x} \leq \vect{x}^TB^{sparse}\vect{x} \leq \vect{x}^T(1+\epsilon)B\vect{x} \quad \forall \vect{x} \in \mathbb{R}^N,
\end{equation}
so that substituting $A = -B$ and rearranging yields the approximation from the main text
\begin{equation}\label{eq:approx_compressed_form}
|\vect{x}^T(A^{sparse} - A)\vect{x}| \leq \epsilon|\vect{x}^TA\vect{x}| \quad \forall \vect{x} \in \mathbb{R}^N,
\end{equation}
where we use the fact that $A$ is negative definite to see that $-\vect{x}^TA\vect{x} = |\vect{x}^TA\vect{x}|$.

By definition, Eq. \ref{eq:approx_compressed_form} approximately preserves $A$ as a quadratic form and thus apart from the eigenvalues and products described below, it also preserves properties of the dynamical system that depend on $A$ as a quadratic form, such as the resting state variances, the diagonal elements of A and the differences-of-covariances (diff-covs).

\textbf{Eigenvalues} Let $\lambda_1 \leq \lambda_2 \leq \dots \leq \lambda_N$ and $\tilde{\lambda}_1 \leq \tilde{\lambda}_2 \leq \dots \leq \tilde{\lambda}_N$ be the eigenvalues of $B$ and $B^{sparse}$ respectively. 

Let $S$ denote the collection of subspaces $U \subset \mathbb R^N$ with $\dim U = k$, and consider the functions $f_B,f_{B^{sparse}}: S \to \mathbb{R}$ given by 
\begin{equation}
f_B(U) = \max_{\substack{\vect{x} \in U \\ \norm{\vect{x}} = 1}} \vect{x}^TB\vect{x}, \qquad f_{B^{sparse}}(U) =  \max_{\substack{\vect{x} \in U \\ \norm{\vect{x}} = 1}} \vect{x}^TB^{sparse}\vect{x}.
\end{equation}
Let $U \in S$ be a given subspace of $\mathbb R^N$ with dimension $k$. Since $(1-\epsilon)\vect{x}^TB\vect{x} \leq \vect{x}^TB^{sparse}\vect{x} \leq (1+\epsilon)\vect{x}^TB\vect{x}$ for all $\vect{x} \in \mathbb{R}^N$, we can take the maximum over all $\vect{x} \in U \subset \mathbb R^N$ with unit norm to see that
\begin{equation}
(1-\epsilon)f_B(U) \leq f_{B^{sparse}}(U) \leq (1+\epsilon) f_B(U).
\end{equation}
Since this inequality holds for any subspace, taking a minimum over all subspaces in $S$ still preserves the inequality:
\begin{equation}
(1-\epsilon)\min_{U \in S}f_B(U) \leq \min_{U \in S} f_{B^{sparse}}(U) \leq (1+\epsilon) \min_{U \in S}f_B(U).
\end{equation}
Thus, by the Courant-Fischer Theorem, we arrive at
\begin{equation}
    (1-\epsilon)\lambda_k \leq \tilde \lambda_k \leq (1+\epsilon)\lambda_k, \quad \forall 1 \leq k \leq N
    \label{eq:ew_preservation}
\end{equation}
Thus all eigenvalues are preserved within a multiplicative factor of $\epsilon$.

\textbf{Eigenvectors} Here we show that the angle between eigenvectors is preserved up to a factor depending on the arbitrarily small degree of spectral approximation $\epsilon$. First, note that a rearrangement of the Davis-Kahan theorem states
\begin{equation}
\sqrt{1 - \frac{4 ||A-A^{sparse}||^2}{\delta_i^2}} \leq \cos \angle(v_i, \tilde v_i), \qquad \text{ for all } i
\end{equation}
where $v_i$ and $\tilde v_i$ are the $i$th eigenvectors of $A$ and $A^{sparse}$ respectively, and 
\begin{equation}
\delta_i = \min_{j: j \neq i} |\lambda_i(A) - \lambda_j(A)| > 0.
\end{equation}

Fix $\gamma > 0$. Now, setting $\epsilon = \frac{\gamma}{2\lambda_{max}}$ and constructing the corresponding $A^{sparse}$, we know
\begin{equation}
||A-A^{sparse}|| = \sup_{||x|| = 1} |x^T(A-A^{sparse})x| \leq \sup_{||x|| = 1} \epsilon |x^TAx| = \epsilon
\lambda_{max} = \frac{\gamma}{2}\end{equation}
where the first equality holds since $A - A^{sparse}$ is Hermitian. Thus, we have
\begin{equation}
\sqrt{1 - \frac{\gamma^2}{\delta_i^2}}  \leq \sqrt{1 - \frac{4 ||A-A^{sparse}||^2}{\delta_i^2}} \leq \cos \angle(v_i, \tilde v_i).
\end{equation}
Since $\gamma > 0$ was arbitrary, this quantity can be made arbitrarily close to $1$. That is, we can guarantee that corresponding eigenvectors of $A$ and $A^{sp}$ point in nearly the same direction.

\textbf{Preserved matrix-vector products} First, note that there exist $N$ linearly independent vectors $\{\vect w_1, \hdots,\vect w_N\}$ (i.e., a basis for $\mathbb{R}^N$) for which the matrix vector products are preserved between $B$ and $B^{sparse}$ (or $A$ and $A^{sparse}$) to within $\epsilon$. Let $\vect{v}_k$ be an eigenvector of $\tilde{I}$ with eigenvalue $1 + \delta_k$ (note that, from Eq. \ref{eq:ew_preservation}, $|\delta_k| \leq \epsilon$). Define $\vect{w}_k = B^{-1/2}\vect{v}_k$ and note that $B\vect{w}_k = B^{1/2}\vect{v}_k$. 
Now 
$B^{sparse}\vect{w}_k = B^{1/2}\tilde{I}B^{1/2}\vect{w}_k = B^{1/2}\tilde{I}\vect{v}_k = (1 + \delta_k)B^{1/2}\vect{v}_k = (1 + \delta_k) B\vect{w}_k$. Consequently, $||(B - B^{sparse})\vect{w}_k|| \leq \epsilon ||B\vect{w}_k||$.

Second, note that the eigenspaces of $B$ and $B^{sparse}$ are close, as we show empirically in Fig. 2 of the main text, though precise bounds will depend on how close the corresponding eigenvalue is to another eigenvalue in the spectrum.

Finally, scalar concentration of measure arguments suggests that a rule of the form in Eq. \ref{eq:sampling_edges} should preserve dense matrix-vector products, provided the entries in the matrix do not grow too large (as for the matrix concentration of measure case). Note, however, that the products of $B$ and $B^{sparse}$ with sparse vectors may be quite different (as will be true for any sparse matrix approximation), because these products are determined by the sum of only a few entries in $B$ and $B^{sparse}$.

\subsection{Partial sampling and robustness to changing probabilities} \label{subsec:partial_samp_and_robustness_to_probs}
\paragraph{Oversampling} The Chernoff bound in Eq. \ref{eq:mat_chernoff}
depends on the sampling probabilities only through the upper bound on the norm of the edge matrices, requiring $0\leq ||Y_{ij}||\leq 1/K_{deg}$. In particular, if the derived $p_{ij}$ for an edge is $<1$ then the same bound holds for any $\tilde{p}_{ij} \geq p_{ij}$ (see below for $p_{ij}>1$). Consequently, the probabilities described in Eq. \ref{eq:prob_B_defn} above are a lower bound, for a given desired degree of approximation ($\epsilon$). If some of the edges are sampled with a greater probability than the theoretical result derived, the approximation equation Eq. \ref{eq:prob_B_defn} will still hold (though some of the terms in the sum of Eq. \ref{eq:id_in_terms_of_B} will have norm less than $1/K_{deg}$). 

The only consequence of over-sampling synapses is that the number of connections in the pruned network will be greater, but the increase is as well behaved as could be desired, corresponding exactly to the degree of over-sampling as $\langle N_{edges} \rangle = \sum_{i>j}p_{ij}$. Furthermore, there is no harm in setting all of the $p_{ii}$ to $1$, as these probabilities do not correspond to edges, but rather the diagonal terms, which relate to the intrinsic leak in the activity of neurons in Eq. \ref{eq:lin_dynamics}. 

Moreover, as long as the sampling probabilities are above the theoretically-derived bound, they can be chosen completely independently at each synapse and do not need to compensate for each other in any way.

\paragraph{Misspecified probabilities} Sampling-based sparsification is quite robust to misspecified sampling probabilities \cite{spielman11}. Again, this robustness emerges because probabilities only affect the Chernoff bound through their effect on the norm of the edge matrices. If some synapses are under-sampled using probability $\hat p_{ij} = \alpha p_{ij}$, with $\alpha<1$, the bound on the $||\tilde Y^{ij}||$'s inflates by a factor of $1/\alpha$ and the degree of approximation becomes $\hat \epsilon = \epsilon/\sqrt{\alpha}$ while preserving the same bound on the probabilities in Eq. \ref{eq:mat_chernoff} (thus maintaining the likelihood of our approximation occuring w.h.p.). To see this, observe that
\begin{equation}
    P\left[\lambda_{min} \leq (1-\hat \epsilon) \right]  \leq N \left(e^{-\hat \epsilon^2/2}\right)^{\alpha/M} = N \left(e^{-\epsilon^2/2}\right)^{1/M},
\end{equation}
and similarly for the other inequality in Eq. \ref{eq:mat_chernoff}. Note here that we could instead choose to maintain our original degree of approximation $\epsilon$, but this would correspond to the larger upper bound 
\begin{equation}
    P\left[\lambda_{min} \leq (1- \epsilon) \right] \leq N \left(e^{-\epsilon^2/2}\right)^{\alpha/M}
\end{equation} 
which means that Eq. \ref{eq:B_psd_ineq} would occur with lower probability.

\paragraph{Fixed edges} The sampling argument can be applied to only a subset of edges in several ways. A particularly natural approach is to simply set the sampling probability for a fixed edge $\tilde{p}_{ij} = 1$ and note that, if  $p_{ij}<1$ the bound $||\tilde Y^{ij}|| \leq M$ still holds and so do the subsequent theoretical results.

A second way to apply the argument to a subset of edges is to write the matrix $A$ as $A_{fixed} + A_{sample}$, where $A_{fixed}$ is the submatrix of edges that are to be preserved and $A_{sample}$ is the submatrix of edges to be either pruned or strengthened. The argument in Section \ref{subsec:sparsification_of_sym_dd_networks} can then be applied to $A_{sample}$ (note that $A_{sample}$ is diagonally dominant and positive semidefinite). This formulation has the disadvantage that the predicted sampling probabilities depend on the covariance matrix determined by $A_{sample}$ rather than $A$, but this covariance matrix may be natural in certain contexts.

Furthermore, while the diagonal terms sampled with probability $p_{ii}$ do not correspond to edges, we can still fix them with no harm to our theoretical results (i.e., set $p_{ii} = 1$ as noted earlier in the oversampling paragraph of this section). 

\paragraph{Synapses with probabilities greater than $1$} A calculated probability term for a synapse $p_{ij}$ that is $>1$ can be handled in two ways. One solution is to convert each synaptic weight into pieces with predicted probability $<1$ and rewrite the sum in Eq. \ref{eq:id_in_terms_of_B} as involving multiple pieces corresponding to the edge $(i,j)$ each sampled with probability $1$. Note that this will increase $K_{deg}$ slightly but does not change the actual form of the sampling rule (the edge is just preserved). A second approach is to split the matrix into a deterministic and a sampled piece, and apply the argument to the sampled piece (as in the argument for fixed edges above). Again this has the drawback that the predicted sampling probabilities would not be given by the covariance matrix of the entire network.

\section{Extensions}
\subsection{Near-diagonally-dominant networks}
Let the matrix $A$ be a (not necessarily diagonally-dominant) symmetric negative definite matrix corresponding to the coupling matrix of a linear system such as in Eq. \ref{eq:lin_dynamics}. We analyze the effect of applying noise-prune to $A$ in terms of its distance from a diagonally dominant matrix.

As before, we let $B = -A$ and analyze the effect of the rule on $B$. Note that the noise-driven covariance matrix of the linear system $C \propto -A^{-1} = B^{-1}$, and that the sampling probabilities yielded by noise-prune are $p_{ij} = K |w_{ij}|(C_{ii} + C_{jj} - 2\sgn(w_{ij})C_{ij}) = K_{deg} |w_{ij}|(B^{-1}_{ii} + B^{-1}_{jj} - 2\sgn(w_{ij})B^{-1}_{ij})$ for $i > j$. We will also set any excess diagonal probabilities $p_{ii} = 1$ (note that this is implicitly done in both the original and matched diagonal settings of noise-prune in the main paper). 

Set $\gamma > 0$ and define the matrix $B_\gamma = B + \gamma I$. Let $B$ have eigenvalues $\lambda_k$ and eigenvectors $\vect{v}_k$ and observe that $B_\gamma$ has eigenvalues $\lambda_i + \gamma$ and the same eigenvectors as $B$. Moreover, note that applying noise-prune to $B$ with some set of probabilities to yield $B^{sparse}$ is equivalent to applying noise-prune to $B_\gamma$ with the same set of probabilities to yield $B_\gamma^{sparse} = B^{sparse} + \gamma I$ (though these are not the optimal probabilities for $B_\gamma$).

Now take $\gamma$ large enough so that $B_\gamma$ is diagonally dominant (the approximation described below will be good if $\gamma$ is small). The framework described in Section \ref{subsec:sparsification_of_sym_dd_networks} can then be applied to $B_\gamma$ and the probabilities saturating the Chernoff bound are $p^{(\gamma)}_{ij}  = K_{deg} |w_{ij}|([B_\gamma]^{-1}_{ii} + [B_\gamma]^{-1}_{jj} - 2\sgn(w_{ij})[B_\gamma]^{-1}_{ij})$. 
 
In particular, note that \begin{equation}
    [B_\gamma]^{-1}_{ij} = \left(\sum_{k} \frac{1}{\lambda_k + \gamma }\vect{v}_k\vect{v}_k^T\right)_{ij} = \sum_{k} \frac{1}{\lambda_k + \gamma}\left(\vect{v}_k\vect{v}_k^T\right)_{ij} = \sum_{k} \frac{1}{\lambda_k + \gamma}(\vect{v}_k)_i(\vect{v}_k)_j \quad \forall i,j
\end{equation} and similarly \begin{equation}
    B^{-1}_{ij}= \left(\sum_{k} \frac{1}{\lambda_k}\vect{v}_k\vect{v}_k^T\right)_{ij} = \sum_{k} \frac{1}{\lambda_k}\left(\vect{v}_k\vect{v}_k^T\right)_{ij} = \sum_{k} \frac{1}{\lambda_k }(\vect{v}_k)_i(\vect{v}_k)_j  \quad \forall i,j,
\end{equation}
where $(\vect{v}_{k})_{\ell}$ denotes the $\ell$th entry of the $k$th eigenvector $\vect{v}_k$. Then we can see that, for $i > j$,
\begin{align*}
    p^{(\gamma)}_{ij} & = K_{deg}|w_{ij}| \sum_k \frac{1}{\lambda_k + \gamma}\left((\vect{v}_k)_i^2  + (\vect{v}_k)_j^2  - 2\sgn(w_{ij})(\vect{v}_k)_i(\vect{v}_k)_j \right) \\
    & = K_{deg}|w_{ij}| \sum_k \frac{1}{\lambda_k + \gamma} ((\vect{v}_k)_i - \sgn(w_{ij})(\vect{v}_k)_j)^2 \\
    & \leq K_{deg}|w_{ij}| \sum_k \frac{1}{\lambda_k} ((\vect{v}_k)_i - \sgn(w_{ij})(\vect{v}_k)_j)^2 \\
    & = p_{ij},
\end{align*}
where the inequality follows from the fact that $\frac{1}{\lambda_k + \gamma} \leq \frac{1}{\lambda_k}$ and the rest of the terms in the expression are all nonnegative.

Thus $p_{ij}^{(\gamma)} \leq p_{ij}$ for all $i \geq j$ and sparsifying $B_{\gamma}$ using the probabilities $p_{ij}$ yields Eq. \ref{eq:B_psd_ineq} for at most the same degree of error $\epsilon$ we would get if we used the $p_{ij}^{(\gamma)}$'s instead (see Section \ref{subsec:partial_samp_and_robustness_to_probs} for more details on oversampling). That is,
\begin{equation}
    (1-\epsilon)\vect{x}^TB_\gamma \vect{x} \leq \vect{x}^T(B_\gamma)^{sparse} \vect{x}\leq (1+\epsilon)\vect{x}^TB_\gamma \vect{x} \quad \forall \vect{x} \in \mathbb{R}^N.
\end{equation}
Now observing that $(B_\gamma)^{sparse} = B^{sparse} + \gamma I$ allows us to subtract $\vect{x}^T \gamma I \vect{x}$ through our inequality to arrive at
\begin{equation}
    (1-\epsilon)x^TB\vect{x} -\epsilon \gamma \vect{x}^T\vect{x}\leq \vect{x}^TB^{sparse}\vect{x} \leq (1+\epsilon)\vect{x}^TB\vect{x} + \epsilon \gamma \vect{x}^T\vect{x} \quad \forall \vect{x} \in \mathbb{R}^N.
\end{equation}
In other words, sparsifying $B$ using the same probabilities we used for $B_{\gamma}$ guarantees a result similar to that of Eq. \ref{eq:B_psd_ineq}, but with an additional additive error of $\epsilon \gamma \vect{x}^T\vect{x}$. In particular, if $\vect{x}$ has unit norm then the additive error is simply $\epsilon \gamma$.

\subsection{Rectified linear units}
Define the rectified linear activation function $[\cdot]_+ = \max[0,\cdot]$ and consider the recurrent neural network
\begin{equation}
    \frac{d\vect{x}}{dt} = - D\vect{x} + [W \vect{x} + \vect{b}(t)]_+.
    \label{eq:relu_dynamics}
\end{equation}
As before, define $A = - D + W$, and let $A^{sparse}$ be the result of applying noise-prune to $A$ using the probabilities from the linear network defined by $A$ (so that Eq. \ref{eq:B_psd_ineq} holds for $A$ and $A^{sparse}$).

Let $\Gamma(t) = \{i: \sum_{j}W_{ij}x_j + b_j(t) > 0\}$ be the indices of neurons that receive suprathreshold input at time $t$. Define $A_{\Gamma(t)}$ and $A_{\Gamma(t)}^{sparse}$ to be the submatrices produced by removing the rows and columns of $A$ and $A^{sparse}$ corresponding to indices not in $\Gamma(t)$. We will show that the dynamics of the network in Eq. \ref{eq:relu_dynamics} are approximately determined by the set of linear systems (indexed by $t$) with coupling matrices $A_{\Gamma(t)}, A_{\Gamma(t)}^{sparse}$. In other words, the dynamics of a rectified linear network switch among the dynamics of a set of linear networks, with the appropriate linear network at a moment in time determined by the subset of neurons that receive suprathreshold input (see \cite{hahnloser00, hahnloser01} for more on this argument).

For convenience, let $\Gamma(t)^c$ be the complement of $\Gamma(t)$; that is, $\Gamma(t)^c$ is the collection of neurons that receive zero input. The neurons in $\Gamma(t)^c$ either have zero activity (and thus can be ignored) or have nonzero activity but receive zero input (and thus contribute feedforward input to the rest of the network that can be absorbed into the input vector). Define $\vect{x}_{\Gamma(t)}$ and $\vect{b}_{\Gamma(t)}$ to be the vectors produced by removing the entries of $\vect{x}$ and $\vect{b}$ corresponding to the indices in $\Gamma(t)^c$, as well as $\vect{x}_{\Gamma(t)^c}$ to be the vector produced by removing the entries of $\vect{x}$ corresponding to the indices in $\Gamma(t)$. Lastly, define $\delta \vect{b}_{\Gamma(t)}$ to be the feedforward contribution of $\vect{x}_{\Gamma(t)^c}$ (more precisely, the $i$th entry of this vector is given by $\sum_{j \in \Gamma(t)^c} W_{ij} x_j$ with $i \in \Gamma(t)$ listed in increasing order) that we will absorb into the new input vector for our smaller system in $\mathbb{R}^{|\Gamma(t)|}$, defined to be $\tilde{\vect{b}}_{\Gamma(t)} = \vect{b}_{\Gamma(t)} + \delta \vect{b}_{\Gamma(t)}$. 

Now the dynamics of the network in some small time interval around $t$ are determined by the linear system,
\begin{align}
\frac{d\vect{x}_{\Gamma(t)}}{dt} =A_{\Gamma(t)} \vect{x}_{\Gamma(t)} + \tilde{\vect{b}}_{\Gamma(t)}. 
\label{eq:relu_dynamics_subsystem}
\end{align}
And the nodes in $\Gamma(t)^c$ either have $0$ activity or are decaying to $0$ with the leak time constant.

Note that Eq. \ref{eq:B_psd_ineq} holds for $A_{\Gamma(t)}, A_{\Gamma(t)}^{sparse}$ as well. Let $\Gamma(t,j)$ be the index of the $j$-th active neuron at time $t$. Given $\vect{x}_{\Gamma(t)} \in \mathbb{R}^{|\Gamma(t)|}$, consider the natural extension vector $\vect{x} \in \mathbb{R}^N$ whose entry in $\Gamma(t, j)$ is the $j$ entry of $\vect{x}_{\Gamma(t)}$ and whose entries in $\Gamma(t)^c$ are $0$. Then $\vect{x}_{\Gamma(t)}^T A_{\Gamma(t)}^{sparse}\vect{x}_{\Gamma(t)} = \vect{x}^TA^{sparse}\vect{x}$ (and similarly, $\vect{x}_{\Gamma(t)}^TA_{\Gamma(t)}\vect{x}_{\Gamma(t)} = \vect{x}^TA\vect{x}$), so the fact that Eq. \ref{eq:B_psd_ineq} holds for $A, A^{sparse}$ implies that it holds for $A_{\Gamma(t)}, A_{\Gamma(t)}^{sparse}$ (for all $t$). Thus, among other quantities, the spectrum of $A_{\Gamma(t)}$ is approximately preserved (to within $\epsilon$) by $A_{\Gamma(t)}^{sparse}$. Thus, we see that noise-prune preserves the dynamics of linear systems described the submatrices $A_{\Gamma(t)}$. Finally, $\tilde{\vect{b}}_{\Gamma(t)}$ depends on the weights through $\delta \vect{b}_{\Gamma(t)}$, which may be perturbed in the sparse system, though it is preserved in expectation. However, perturbations are likely to be small because this additional feedforward input comes from the small subset of low-activity neurons in $\vect{x}_{\Gamma(t)^c}$ that receive sub-threshold input and are approaching zero activity but have not completely decayed yet (which they do so with time-constant given by the leak). In short, the dynamics of a rectified linear network are approximately preserved when its coupling matrix is sparsified in the same manner as that of a linear network. 

\bibliography{../noisy_pruning}